\documentclass[journal]{IEEEtran}
\usepackage{cite}

\ifCLASSINFOpdf
  \usepackage[pdftex]{graphicx}
  \graphicspath{{./graphics/}}
  \DeclareGraphicsExtensions{.pdf,.jpeg,.png}
\else
\fi
\usepackage{amsmath}
\usepackage{array}
\usepackage{tabularx}

\ifCLASSOPTIONcompsoc
 \usepackage[caption=false,font=normalsize,labelfont=sf,textfont=sf]{subfig}
\else
 \usepackage[caption=false,font=footnotesize]{subfig}
\fi
\usepackage{url}

\usepackage{color,soul,colortbl}
\definecolor{lightred}{RGB}{255,179,186}
\definecolor{lightgreen}{RGB}{186,255,201}
\definecolor{lightblue}{RGB}{186,225,255}
\definecolor{lightpurple}{RGB}{192,174,231}
\definecolor{lightorange}{RGB}{255,223,186}
\definecolor{lightpink}{RGB}{255,181,232}
\definecolor{lightgrey}{RGB}{192,192,192}

\begin{document}
%
\title{Attacking the Nintendo 3DS Boot~ROMs}
%
%
%

\author{Michael~Scire,
        Melissa~Mears,
        Devon~Maloney,
        Matthew~Norman,
        Shaun Tux,
        and Phoebe Monroe \vspace{-2.0em}}

%
%

\maketitle

\begin{abstract}
We demonstrate attacks on the boot~ROMs of the Nintendo 3DS in order to
exfiltrate secret information from normally protected areas of memory and gain
persistent early code execution on devices which have not previously been
compromised. The attack utilizes flaws in the RSA signature verification
implementation of one of the boot~ROMs in order to overflow ASN.1 length fields
and cause invalid firmware images to appear valid to the signature parser. This
is then used to load a custom firmware image which overwrites the data-abort
vector with a custom data abort handler, then induces a data-abort exception in
order to reliably redirect boot~ROM code flow at boot time. This executes a
payload which, due to its reliable early execution by a privileged processor, is
able to function as a persistent exploit of the system in order to exfiltrate
secret information (such as encryption keys) from normally protected areas of
memory.
\end{abstract}


\section{Introduction}

As with all entertainment consoles, the Nintendo 3DS has the difficult task of
accommodating legitimate users while still enforcing a reasonable level of
device security. In pursuit of accomplishing this, all models of the Nintendo
3DS are equipped with two separate processors located in a single System-on-Chip
(``SoC''). There is a group ARM11 application processor (``ARM11'') which runs
the OS and is responsible for userspace-level tasks, and a group ARM9 security
processor (``ARM9'') which is responsible for enforcing permission rights and
managing access to cryptographic hardware and the filesystem. This system of
multiple processors with varying levels of access, designed with the principle
of least privilege in mind, allows for a chain of trust which ensures only
authorized code is executed \cite{cryptosystem}.

The root of trust for 3DS security, as with most embedded systems, is found
within the boot~ROMs burned into the processors at the factory. Specifically,
the ARM9 boot~ROM (``Boot9'') contains RSA public keys to which only Nintendo
has the matching private keys. This is used to implement a secure bootchain
designed to ensure that only firmware images signed by Nintendo will run on the
system. This bootchain, rooted at the public keys in Boot9, is responsible for
initializing device hardware and securely launching system firmware from
storage\cite{bootloader}.

To help facilitate the security of this bootchain, both Boot9 and its ARM11
counterpart (``Boot11'') are split into two halves. The first half is readable
from firmware given sufficient privileges, while the second half is made
unreadable early in the boot process by setting specific registers which can
only be written to once per boot. By combining this concept of ``unprotected''
and ``protected'' halves with implementations of AES and RSA write-only
``keyslots'' (secure memory areas readable only by the respective hardware AES
and RSA implementations), Nintendo is able to store and use secret keys in such
a way that even a fully compromised system will not expose all secure data
\cite{AES_Registers}\cite{RSA_Registers}.

\section{Secure Bootchain}

\subsection{Known Information}

In the ``Nintendo Hacking 2016'' talk given at the 33\textsuperscript{rd} Chaos
Communication Congress (``33C3''), one of the speakers (``derrek'') detailed the
use of a known hardware ``Vector-Glitch'' attack to dump the protected half of
Boot9. This approach was reported to be highly unstable and, due to Boot9 being
unshareable copyrighted code, would have to be re-implemented by any successive
attacker (requiring access to prohibitively expensive hardware). Although this
means that the speaker's Vector-Glitch attack is impractical to reproduce
consistently, the talk is still useful due to the information it provides about
the protected half of Boot9\cite{33c3}.

\subsection{Firmware Images}

The firmware image loaded by Boot9 is stored encrypted on the boot device. The
plaintext of this encrypted image consists of a firmware image header and up to
four firmware `sections'. The firmware image header contains a magic value
(\texttt{FIRM}), a boot priority, the ARM9 and ARM11 entrypoints, the firmware
section headers, and an RSA-2048 signature over a SHA-256 hash of the rest of
the firmware header. The firmware section headers each contain the offset of the
firmware section, the physical address to load the section to, the size of the
firmware section, the copy method to use in loading the firmware section, and a
SHA-256 hash of the firmware section. In a stock firmware image, the firmware
sections contain the ARM9 kernel, the ARM11 kernel, and other data required to
initialize the system\cite{FIRM}.

\subsection{Boot Process}

All 3DS models have identical boot~ROMs burned in at the factory, and thus have
the same initial boot process implementation which is performed by
Boot9\cite{bootloader}\cite{33c3}:

\medskip
\begin{enumerate}
  \item Initialize AES keyslots with secret keys from Boot9 (using AES hardware
  to decrypt all subsequent data read from boot device)
  \item Initialize RSA keyslots with firmware public keys from Boot9 (using
  separate keys for various signature and console types)
  \item Select boot device (typically NAND flash storage)
  \item Read firmware header from selected boot device to memory
  \item Validate SHA-256 hash and RSA-2048 signature of firmware header
  \item Read firmware sections to memory according to parameters specified by
  firmware section headers
  \item Disable access to protected halves of Boot9 and Boot11
  \item Jump to ARM9 and ARM11 entrypoints specified by firmware header
\end{enumerate}
\medskip

This is a simple, robust boot process which seems secure from a conceptual
perspective. Any flaws in this bootchain must, therefore, lie in the
implementation rather than the theory.

\section{Signature Verification}

\subsection{Known Information}

In the previously mentioned ``Nintendo Hacking 2016'' talk, the existence of
several flaws in the implementation of this seemingly secure boot process were
revealed\cite{33c3}. Unfortunately, performing an attack based on these flaws
requires details which can only be found by examining the protected half of
Boot9. The speakers of the talk kept these details a secret, which required us
to pursue a more creative approach to exploit these flaws.

\subsection{Firmware Header Signatures}

Recall that RSA signatures are, essentially (as an oversimplification), a
specifically formatted hash which has been encrypted by a signer's private key.
These signatures are then verified by calculating the hash of the data,
decrypting the the signature using the signer's matching public key, then
verifying the hash which was calculated matches the hash embedded in the
signature. The exact formatting and `padding' used for the hash in an RSA
signature are extremely important for preventing numerous attacks on it.

In the case of the 3DS, the private key is possessed only by Nintendo, the
public key is embedded in Boot9, and the data being verified is the firmware
header loaded by Boot9. Specifically, the firmware header is signed by a
PKCS\#1v1.5 padded RSA-2048 signature with an embedded ASN.1 encoded SHA-256
hash. An RSA-2048 signature is \texttt{0x100} bytes in length, while a SHA-256
hash is only \texttt{0x20} bytes in length. This means that the remaining bytes
are filled by a deterministic padding of \texttt{FF} bytes followed by a fixed
ASN.1 DER encoding of the hash type.

To parse this relatively complex structure, Nintendo decided to write their own
signature parser. This signature parser is flawed in the following
ways\cite{33c3}:

\medskip
\begin{enumerate}
  \item Bounds checking is not performed when parsing one or more ASN.1 length
  fields, allowing for length values to point beyond the signature block
  \item PKCS\#1v1.5 Block Type 2 (designed for encrypted messages, not
  signatures) is permitted, allowing for arbitrary padding rather than just
  \texttt{FF} bytes
  \item Padding is not required to completely fill the signature block, allowing
  for a less strict signature layout
\end{enumerate}
\medskip

The first flaw is critical because the parser adds the ASN.1 length fields to
the current memory offset to arrive at the location of the SHA-256 hash embedded
in the signature (``embedded hash'') used for comparison. This means a signature
can be calculated where the length fields overflow valid lengths and cause the
parser to look elsewhere on the stack for the embedded hash.

This allows for an attack in which very specific length fields cause the
signature parser to use the exact area of memory on the stack where it already
stored the calculated SHA-256 hash of the firmware header (``calculated hash''),
rather than the area of memory on the stack where the embedded hash is. This
results in the signature parser comparing the calculated hash with itself
instead of the embedded hash, which never fails. This means any firmware image
using this signature will appear to be validly signed to the signature parser,
enabling the ``fakesigning'' of arbitrary firmware images.

The second and third flaws severely weaken the security of the signature itself
because they allow for attacks on RSA which would normally be stopped by strict
padding checks. This means that a search for a valid signature to exploit the
first flaw can take place in a significantly shorter amount of time than would
otherwise be possible.

\section{Blind Exploitation}

\subsection{Missing information}

As mentioned previously, knowledge of these flaws alone is not sufficient to
perform an attack. This is because the critical flaw allows for the the use of
crafted length fields to modify where the parser looks for the embedded hash on
the stack, which is only useful if the parser can be made to look at the
calculated hash. This requires knowledge of both which length fields are
improperly verified and where in the stack the calculated signature is located.
The first approach to try is to search for possible re-use of the flawed
signature parser elsewhere.

Examining the signature parser used by the firmware itself to verify the
integrity of software reveals it to be very similar to the described Boot9
signature parser, though with the addition of strict padding checks which
prevent an attack. These padding checks are present in all firmware versions
from 1.0.0 (kernel 0.14) onward, meaning that these flaws were noticed and fixed
some time after Boot9 was finalized but before firmware 1.0.0 was shipped.

\subsection{Factory Firmware}

NAND flash storage, as with most non-volatile storage mediums, does not actually
erase information completely when instructed to (in most cases); rather,
information is simply marked as erased and can be overwritten by new information
at a later time. By examining a decrypted 3DS NAND flash image, it is possible
to recover fragments of data left over from the firmware installed to the device
at the factory (which was erased from the device before it was sold).

When enough devices are examined (especially those which have only been lighly
used), it is possible to re-assemble these fragments into a pre-1.0.0 'factory
firmware' (kernel 0.13). When this factory firmware is examined, we see a
similar signature parser to firmware 1.0.0, but without the strict padding
checks. This means that factory firmware is vulnerable to the same signature
verification flaws as Boot9, and thus its signature parser can be inferred to be
very similar to the signature parser used in Boot9.

\subsection{Signature Calculations}

Note that different public keys are used for the various signature and console
types (a total of 6). These are split into the `Retail' and `Developer'
categories, where the former is used in standard consoles and the latter is used
in official Nintendo Developer Program Consoles.

The three signature types within these two categories are the following: the
`NCSD header' signature (verifies the partition layout and encryption methods),
the `NAND boot' signature (verifies firmware images loaded from NAND flash
storage), and the `non-NAND boot' signature (verifies firmware images loaded
from sources other than NAND flash storage). For this initial test, a retail
device booting from NAND flash storage was chosen.

By examining the signature parser used in factory firmware, we can both
determine which ASN.1 length fields are incorrectly verified and the location on
the stack where the calculated hash will be. With this information, we then
calculated a signature which correctly overflows the ASN.1 length fields to
exploit the factor firmware signature parser (``exploit signature'') and tested
it on Boot9. Unfortunately, this led us to the discovery that, while the
signature parser used by factory firmware is very similar to the signature
parser used in Boot9, they are not identical.

Interestingly, this caused Boot9 to display a black screen rather than the
standard blue error screen normally shown when Boot9 encounters an error (such
as a failed signature check). This means that, while the signature was
successful in causing the parser to look elsewhere on the stack for the embedded
hash, the memory the parser was trying to access was invalid which was resulting
in an unhandled data-abort exception. From this, we know that the calculated
hash is in a different location on the stack than it is for factory firmware.

\subsection{Brute-forcing Signatures}

We can calculate that, for this signature parser, there are only 128 possible
locations on the stack (relative to the signature) for which a signature can be
brute-forced in a reasonable amount of time to exploit the length field
vulnerability. Because this exploit has been performed before, we know that the
calculated hash for Boot9's signature parser must be in one of these 128
locations. We can then search through randomly generated signatures (a
``brute-force'') to find valid signatures that cause the signature parser to
look for the embedded hash at these 128 locations.

Recall that RSA signatures are created by powering plaintext message \(m\) by
secret \(d\) using modulus \(n\), and verified by powering signature \(s\) by
RSA exponent \(e\) (65537 in this case) using modulus \(n\):

\[s \equiv m^{d} \pmod{n} \quad \text{and} \quad v \equiv s^{e} \pmod{n}\]

The simple method for calculating signature \(s\) for an exploit plaintext
\(m\) would be to calculate \(m \equiv s^{65537} \pmod{n}\) for a random \(s\)
until a usable \(m\) is found. However, this method is extremely slow because it
requires 17 multiplication operations and 17 modulo operations per attempt.

One way to improve the speed of this signature search is to check \(-m\) when
checking \(m\). This is because, if \(-m\) is a match, then \(-s\) is its
signature. Because subtraction is significantly faster than multiplication, this
method nearly doubles the speed of our search.

Another way to improve the speed of our search is to take advantage of the
second and third flaw in the padding checks of Boot9's signature parser. Because
we do not have to adhere to strict padding checks, an `RSA Multiplicative
Attack' is possible: if \(s_{a}\) and \(s_{b}\) are signatures for messages
\(a\) and \(b\), then \(s_{a}s_{b} \pmod{n}\) is a signature for message \(ab
\pmod{n}\). This results in another significant performance gain. This results
in the following brute-force search method:

\medskip
\begin{enumerate}
  \item Select random root \(r\) such that \(1 < r < n\)
  \item Calculate multiplier \(k\) such that \(k = r^{65537} \pmod{n}\)
  \item Set \(y = 1\) and \(z = 0\)
  \item Define parameters for validity (based on 128 possible locations)
  \item Loop:
    \begin{enumerate}
      \item Increment \(z\) by 1
      \item Perform \(y = (y * k) \pmod{n}\)
      \item Check whether \(y\) is valid
        \begin{itemize}
          \item Return \(r^{z} \pmod{n}\)
        \end{itemize}
      \item Check whether \(n - y\) is valid
        \begin{itemize}
          \item Return \(-r^{z} \pmod{n}\)
        \end{itemize}
    \end{enumerate}
\end{enumerate}
\medskip

While it is difficult to express the exact probability of finding a valid
signature with this method due to the complexities in representing the
parameters of a valid signature mathematically, we hypothesized that the
probability of finding a valid signature was approximately 1 in \(2^{43}\).

We then used a modular multiplication algorithm for Graphics Processing Units
(``GPUs'') created by Kaiyong Zhao\cite{modmult} to write an implementation of
this brute-force search method that could find valid RSA signatures for each
signature and console type\cite{sighaxgpu}. This implementation was then run
using a few desktop machines with high-end consumer GPUs and two Amazon Web
Services GPU compute (``p2.8xlarge'') instances.

By trying each valid signature as it was found, we determined that the correct
relative location of the calculated hash for the Boot9 signature parser is
immediately after the signature on the stack.

\begin{figure}[h]
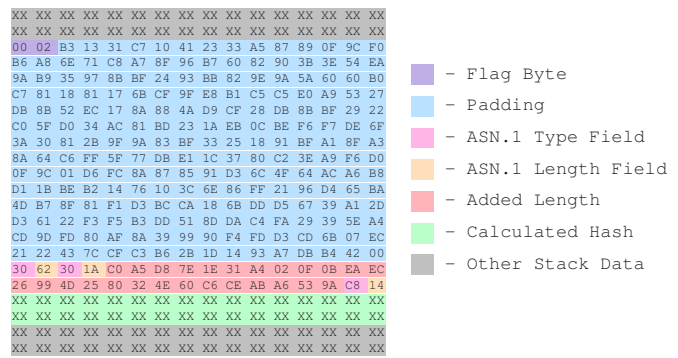

  \begin{minipage}{.60\linewidth}
    \tiny{\texttt{\noindent
      \sethlcolor{lightgrey}\hl{XX XX XX XX XX XX XX XX XX XX XX XX XX XX XX XX\\
      XX XX XX XX XX XX XX XX XX XX XX XX XX XX XX XX}\\
      \sethlcolor{lightpurple}\hl{00 02 }\sethlcolor{lightblue}\hl{B3 13 31 C7 10 41 23 33 A5 87 89 0F 9C F0\\
      B6 A8 6E 71 C8 A7 8F 96 B7 60 82 90 3B 3E 54 EA\\
      9A B9 35 97 8B BF 24 93 BB 82 9E 9A 5A 60 60 B0\\
      C7 81 18 81 17 6B CF 9F E8 B1 C5 C5 E0 A9 53 27\\
      DB 8B 52 EC 17 8A 88 4A D9 CF 28 DB 8B BF 29 22\\
      C0 5F D0 34 AC 81 BD 23 1A EB 0C BE F6 F7 DE 6F\\
      3A 30 81 2B 9F 9A 83 BF 33 25 18 91 BF A1 8F A3\\
      8A 64 C6 FF 5F 77 DB E1 1C 37 80 C2 3E A9 F6 D0\\
      0F 9C 01 D6 FC 8A 87 85 91 D3 6C 4F 64 AC A6 B8\\
      D1 1B BE B2 14 76 10 3C 6E 86 FF 21 96 D4 65 BA\\
      4D B7 8F 81 F1 D3 BC CA 18 6B DD D5 67 39 A1 2D\\
      D3 61 22 F3 F5 B3 DD 51 8D DA C4 FA 29 39 5E A4\\
      CD 9D FD 80 AF 8A 39 99 90 F4 FD D3 CD 6B 07 EC\\
      21 22 43 7C CF C3 B6 2B 1D 14 93 A7 DB B4 42 00}\\
      \sethlcolor{lightpink}\hl{30 }\sethlcolor{lightorange}\hl{62 }\sethlcolor{lightpink}\hl{30 }\sethlcolor{lightorange}\hl{1A }\sethlcolor{lightred}\hl{C0 A5 D8 7E 1E 31 A4 02 0F 0B EA EC\\
      26 99 4D 25 80 32 4E 60 C6 CE AB A6 53 9A }\sethlcolor{lightpink}\hl{C8 }\sethlcolor{lightorange}\hl{14}\\
      \sethlcolor{lightgreen}\hl{XX XX XX XX XX XX XX XX XX XX XX XX XX XX XX XX\\
      XX XX XX XX XX XX XX XX XX XX XX XX XX XX XX XX}\\
      \sethlcolor{lightgrey}\hl{XX XX XX XX XX XX XX XX XX XX XX XX XX XX XX XX\\
      XX XX XX XX XX XX XX XX XX XX XX XX XX XX XX XX}
    }}
  \end{minipage}%
  \begin{minipage}{.40\linewidth}
    \linespread{1.3}{\scriptsize{\texttt{\noindent
      \sethlcolor{lightpurple}\hl{ {} } - Flag Byte\\
      \sethlcolor{lightblue}\hl{ {} } - Padding\\
      \sethlcolor{lightpink}\hl{ {} } - ASN.1 Type Field\\
      \sethlcolor{lightorange}\hl{ {} } - ASN.1 Length Field\\
      \sethlcolor{lightred}\hl{ {} } - Added Length\\
      \sethlcolor{lightgreen}\hl{ {} } - Calculated Hash\\
      \sethlcolor{lightgrey}\hl{ {} } - Other Stack Data\\
    }}}
  \end{minipage}%
  \caption{Structure of a plaintext exploit signature in memory}
\end{figure}

With our search method, we found 1 valid signature per approximately 8 trillion
attempts, which was within 10\% of our estimate (though we do acknowledge that
our sample size was small). At this rate, on this hardware, it took
approximately 9 days to find valid signatures for all 6 signature variants.

With these signatures, we now have the capability to fakesign custom firmware
images and run our own code. Additionally, because the Boot9 is burned into the
ARM9 processor, this RSA signature parsing vulnerability cannot be patched with
an update to the system firmware.

\section{Exfiltrating Secrets}

\subsection{Utilizing New Privileges}

Unfortunately, because access to the protected boot~ROMs is disabled in the boot
process just before Boot9 jumps to the firmware entrypoint, merely having the
ability to run custom code from fakesigned firmware images is not enough to gain
access to the protected halves of Boot9 and Boot11. Fortunately, this is not the
only ability we have gained. 

Recall that the various firmware header and firmware section header parameters
include, among other things, the boot priority, the ARM9 and ARM11 entrypoints,
the physical addresses to load firmware sections to, the size of firmware
sections, and the copy method used in loading the firmware section. Of these,
the most interesting is the location the physical addresses firmware sections
are loaded to.

\subsection{Firmware Section Copying}

In the previously mentioned ``Nintendo Hacking 2016'' talk, we learned that
Boot9 checks the location specified in the firmware section header against a
blacklist of memory regions, but they are not very thorough. Specifically, only
the Boot9 data regions are blacklisted. Fortunately, there are other data
regions which are just as dangerous.

\begin{figure}[h]
  \footnotesize{\texttt{
    \setlength\extrarowheight{1.5pt}
    \begin{tabularx}{\linewidth}{lrrl}
      Region & Address    & Size       & Description \smallskip\\
      0      & 0x20000000 & 0x08000000 & FCRAM \\\rowcolor{lightgreen}
      1      & 0x10000000 & 0x10000000 & I/O Registers \\
      2      & 0x08000000 & 0x00100000 & ARM9 Memory \\
      3      & 0x08000000 & 0x00000400 & ARM9 Memory \\
      4      & 0xFFF00000 & 0x00004000 & DTCM \\
      5      & 0x07FF8000 & 0x00008000 & ITCM \\
      6      & 0xFFFF0000 & 0x00010000 & Boot9 Data \\
      7      & 0x1FFFE000 & 0x00000800 & AXI WRAM \\
    \end{tabularx}
  }}
  \caption{ARM9 memory layout \cite{memorylayout}}
\end{figure}

The area of immediate interest is the I/O registers. Specifically, the `New'
Direct Memory Access (``NDMA'') engine is a memory-mapped I/O register. By
constructing an NDMA copy request as a firmware section and loading it to the
NDMA registers using the firmware section header, the NDMA engine will perform a
Direct Memory Access (``DMA'') to copy the protected half of Boot9 to somewhere
else in memory.

Because this NDMA copy is triggered by the loading of the firmware section, it
occurs before access to Boot9 is disabled. We gave this copy request a firmware
section header which loaded it to the NDMA registers, then placed in a
fakesigned firmware image which would boot normally after performing the
request. We were then able to successfully retreive the protected half of Boot9
from ARM9 memory after the boot completed.

\section{Code Execution}

\subsection{Boot9 Code Execution}

By using the NDMA engine in the way described, we can now copy data from our
firm sections to any location. Of these locations, one of the most interesting
is the exception vectors. Because there are no restrictions on where we can copy
data, this allows for an attack in which we overwrite the data-abort vector with
our own handler before inducing a data-abort by attempting to copy to
\texttt{NULL}. This allows us to redirect code flow to gain Boot9 code execution
before access to the protected half of Boot9 is disabled.

\subsection{Boot11 Code Execution}

Though we have managed to thoroughly break the security of Boot9, this does not
yet allow for access to the protected half of Boot11. Fortunately, this task is
made relatively easy by the access we have gained up to this point. The ARM11
processor operates in a \texttt{0x80000} byte region of `work' SRAM (``WRAM'')
connected to the SoC by Advanced Extensible Interface (``AXI'').

Because we already have reliable Boot9 code execution, we have full access to
the ARM11's memory from Boot9. This allows us to trivially overwrite a specific
Boot11 function pointer in AXI WRAM at the correct time, replacing it with a
pointer of our own. When this pointer is dereferenced by Boot11, code flow will
be redirected and we will gain Boot11 code execution before the protected half
of Boot11 is disabled, allowing us to gain access to the protected half of
Boot11.

\subsection{Implementation Details}

It is important to note that access to the Fast Cycle DRAM (``FCRAM'') is
enabled for each processsor by the same register which disables access to the
protected half of that processor's boot~ROM. Because access to FCRAM is
necessary for running the 3DS OS, it is not sufficient to simply skip enabling
these registers and leave the boot~ROMs unprotected. This means that the
implementation of these code execution vulnerabilities must take this into
account.

For this implementation, we use a fakesigned custom firmware image comprised of
4 firmware sections. Section 0 contains a Boot11 hook and an ARM11 `Stage 2'
payload, which are copied to AXI WRAM. Section 1 contains a custom data-abort
handler, a custom data-abort vector, two Boot9 hooks, and an ARM9 `Stage 2'
payload, which are copied to a safe area in ARM9 memory. Section 2 contains NDMA
copy reqests to copy the data from Section 1 to the correct locations, which is
copied to the NDMA registers. Section 3 contains invalid data, which is loaded
to \texttt{NULL} to induce a data-abort.

After all of these firmware sections are copied to the correct locations, the
attempt to copy the invalid data of Section 3 triggers a data-abort exception.
When this happens, code flow is redirected to our installed data-abort handler
which installs both Boot9 hooks by overwriting specific Boot9 function pointers.
The data-abort handler then skips the copy instruction which caused the
data-abort before continuing the normal ARM9 boot process. 

When Boot9 dereferences the first overwritten function pointer, the first Boot9
hook is called. This hook installs the Boot11 hook by overwriting a
specific Boot11 function pointer, sets up the Memory Protection unit (``MPU''),
and sets a specific flag in memory to communicate with Boot11. It then continues
the normal ARM9 boot process.

When Boot9 dereferences the second overwritten function pointer, the second
Boot9 hook is called. This hook will wait in a loop until it receives a signal
from the Boot11 hook. While this is happening, the Boot11 hook is called at some
point in the normal ARM11 boot process. This hook copies the protected half of
Boot11 to a specific location in AXI WRAM, signals to the second Boot9 hook,
then waits in a loop until it receives a signal from the second Boot9 hook.

Once the second Boot9 hook receives the signal from the Boot11 hook, it copies
the protected half of Boot11 to a safe location in ARM9 memory, signals to the
Boot11 hook, then copies the protected half of Boot9 to another safe location in
ARM9 memory. At this point, both protected boot~ROM halves have been copied to
ARM9 memory and both processors are synchronized.

Once the hooks are finished, both processors then jump to the Stage 2
entrypoints. Stage 2 will check if a specific key combination is being held
down. If it is, it will copy the protected halves of Boot9 and Boot11 to the
device's SD card and shut down. Otherwise, it will attempt to load a second
custom firmware image from the SD card, activate access to FCRAM by disabling
access to the protected boot~ROMs (if requested to by the custom firmware
image), then continue the boot process. This second custom firmware image is
user-replacable and allows for adding in-memory patches to the official device
firmware on boot.

\subsection{Boot Devices}

Recall that Boot9 has three public key signature types: the NCSD header
signature, the NAND boot signature, and the non-NAND boot signature. We have
already explored how the ability to create NAND boot signatures which appear
valid can be useful to an attacker who already has sufficient device access to
install a firmware image to NAND flash storage, but it is difficult to reach
that level of access because a chain of exploits is needed (the firmware's
signature parser is not flawed like Boot9's). Fortunately, we have the ability
to create valid signatures for each of the other types too.

The NCSD header signature type is not particularly interesting because it also
requires access to NAND flash storage, but the remaining non-NAND boot signature
type sounds promising. In the previously mentioned ``Nintendo Hacking 2016''
talk, the ability for Boot9 to load signed firmware images from the Wi-Fi SPI
flash storage was mentioned (presumably for device repair purposes), but an
examination of this storage shows it to be unwritable without special hardware,
and is thus even less accessible than NAND flash storage for booting a custom
firmware image.

What was not mentioned, however, was the second alternative boot method we
discovered upon examining the protected half of Boot9. Before Boot9 attempts to
load a firmware image from NAND, it checks to see if the device's shell is
closed and if a specific key combination (\texttt{START} + \texttt{SELECT} +
\texttt{X}) is being held. If both of these are done, Boot9 will check if a
standard DS cartridge (``NTR'' cartridge) is inserted, then attempt to load a
signed firmware image from it.

By utilizing an exploit RSA signature for the non-NAND boot signature type, it
is possible to boot a custom firmware image from one of the many commonly
available rewritable NTR cartridges (``flashcarts''). We can then use a similar
implementation as before, but this time have Stage 2 utilize its ARM9 code
execution to install our original custom firmware image implementation to NAND
flash storage.

This flashcart is then inserted into the device. Though it should be impossible
to press the required key combination while the shell is closed due to the
device's design, a magnet can be used to trick the shell sensor. We then just
hold the required key combination and power on the device, which loads our
custom firmware image and installs the previously described exploit
implementation to NAND flash storage.

\section{Conclusion}

We have demonstrated attacks on the secure bootchain of the Nintendo 3DS which
defeat its primary protection mechanisms. We used `blind exploitation'
techniques to forge signatures and exfiltrate device secrets based on limited
information, then took advantage of weak internal security mechanisms to gain
reliable early code execution and a method to exploit devices which have not
previously been compromised. This weak internal security was best demonstrated
by the failure to properly validate the copy locations specified in firmware
section headers, showing an implicit trust placed in the signature validation
process. This use of signed malicious firmware images shows the need for strong
internal security mechanisms to mitigate damage in the case of primary security
mechanisms being defeated.



\bibliographystyle{IEEEtran.bst}
\bibliography{IEEEabrv,references}

\begin{thebibliography}{1}
\providecommand{\url}[1]{#1}
\csname url@samestyle\endcsname
\providecommand{\newblock}{\relax}
\providecommand{\bibinfo}[2]{#2}
\providecommand{\BIBentrySTDinterwordspacing}{\spaceskip=0pt\relax}
\providecommand{\BIBentryALTinterwordstretchfactor}{4}
\providecommand{\BIBentryALTinterwordspacing}{\spaceskip=\fontdimen2\font plus
\BIBentryALTinterwordstretchfactor\fontdimen3\font minus
  \fontdimen4\font\relax}
\providecommand{\BIBforeignlanguage}[2]{{%
\expandafter\ifx\csname l@#1\endcsname\relax
\typeout{** WARNING: IEEEtran.bst: No hyphenation pattern has been}%
\typeout{** loaded for the language `#1'. Using the pattern for}%
\typeout{** the default language instead.}%
\else
\language=\csname l@#1\endcsname
\fi
#2}}
\providecommand{\BIBdecl}{\relax}
\BIBdecl

\bibitem{cryptosystem}
\BIBentryALTinterwordspacing
Y.~Lu. (2016, apr) The 3ds cryptosystem. [Online]. Available:
  \url{https://yifan.lu/2016/04/06/the-3ds-cryptosystem/}
\BIBentrySTDinterwordspacing

\bibitem{bootloader}
\BIBentryALTinterwordspacing
3dbrew Contributors. Bootloader. [Online]. Available:
  \url{https://www.3dbrew.org/wiki/Bootloader}
\BIBentrySTDinterwordspacing

\bibitem{AES_Registers}
\BIBentryALTinterwordspacing
------. Aes registers. [Online]. Available:
  \url{https://www.3dbrew.org/wiki/AES_Registers}
\BIBentrySTDinterwordspacing

\bibitem{RSA_Registers}
\BIBentryALTinterwordspacing
------. Rsa registers. [Online]. Available:
  \url{https://www.3dbrew.org/wiki/RSA_Registers}
\BIBentrySTDinterwordspacing

\bibitem{33c3}
\BIBentryALTinterwordspacing
derrek, nedwill, and naehrwert. (2016, dec) Nintendo hacking 2016. [Online].
  Available: \url{https://media.ccc.de/v/33c3-8344-nintendo_hacking_2016}
\BIBentrySTDinterwordspacing

\bibitem{FIRM}
\BIBentryALTinterwordspacing
3dbrew Contributors. Firm. [Online]. Available:
  \url{https://www.3dbrew.org/wiki/FIRM}
\BIBentrySTDinterwordspacing

\bibitem{modmult}
\BIBentryALTinterwordspacing
K.~Zhao, ``Implementation of multiple-precision modular multiplication on
  gpu,'' 2009. [Online]. Available:
  \url{https://www.comp.hkbu.edu.hk/\~pgday/2009/10th_papers/kzhao.pdf}
\BIBentrySTDinterwordspacing

\bibitem{sighaxgpu}
\BIBentryALTinterwordspacing
Myria. sighaxgpu. [Online]. Available:
  \url{https://github.com/Myriachan/sighax}
\BIBentrySTDinterwordspacing

\bibitem{memorylayout}
\BIBentryALTinterwordspacing
3dbrew Contributors. Memory layout. [Online]. Available:
  \url{https://www.3dbrew.org/wiki/Memory_layout}
\BIBentrySTDinterwordspacing

\end{thebibliography}
%



\end{document}